\documentclass[aps,prl,twocolumn,showpacs]{revtex4-1}

\usepackage{graphicx}
\usepackage{color}
\usepackage{amsmath}
\usepackage{amssymb}
\usepackage[normalem]{ulem}
\usepackage[usenames,dvipsnames]{xcolor}

\begin{document}

\title{Quench-induced breathing mode of one-dimensional Bose gases}
\date{\today}
\author{Bess Fang, Giuseppe Carleo, and Isabelle Bouchoule}
\affiliation{Laboratoire Charles Fabry, Institut d'Optique,
  Univ Paris Sud 11, 2 avenue Augustin Fresnel, F-91127 Palaiseau
  cedex, France}

\begin{abstract}
We measure the position- and momentum-space breathing dynamics of
trapped one-dimensional Bose gases at finite temperature. The profile
in real space reveals sinusoidal width oscillations whose frequency
varies continuously through the quasicondensate to ideal Bose gas
crossover. A comparison with theoretical models taking temperature
into account is provided.  In momentum space, we report the first
observation of a frequency doubling in the quasicondensate regime,
corresponding to a self-reflection mechanism due to the repulsive
interactions.  
Its disappearance
through the crossover is mapped out experimentally, giving insights to
the dynamics of the breathing evolution.
\end{abstract}

\pacs{03.75.Kk, 67.85.-d}

\maketitle 

The field of ultracold atomic and molecular gases has gained
increasing significance in the study of nonequilibrium phenomena in
quantum many-body systems.  The accurate time-dependent control of
microscopic parameters enables the realization of prototypical
nonequilibrium processes, complementary to those studied in condensed
matter physics.  In particular, isolated quantum many-body systems can
be driven out of equilibrium and accurately monitored.  The
realization of such experimental
simulators~\cite{trotzky_probing_2012} therefore constitutes a unique
tool to investigate fundamental questions. Remarkable examples are the
observation of a light-cone effect in the spreading of
correlations~\cite{langen_local_2013, cheneau_light-cone-like_2012},
as well as the studies of quantum
ergodicity~\cite{trotzky_probing_2012, kinoshita_quantum_2006,
  gring_relaxation_2012, ronzheimer_expansion_2013}, where
prethermalisation was observed in (nearly) integrable systems.

One-dimensional (1D) systems are ideal test benches for the study of
out-of-equilibrium phenomena because of their intrinsic strong
correlations and the possibility to realize integrable models.  The
experimental investigation of their quantum dynamics is of particular
relevance in clarifying questions for which a theoretical treatment is
challenging.  The understanding of the time-dependent behavior of
interacting quantum systems is indeed restricted to a few limiting
cases~\cite{polkovnikov_colloquium:_2011}, and plagued by the lack of
systematic \emph{ab initio} approaches.  This stands in 
contrast to the 
well consolidated understanding of
thermodynamic properties of such systems, for which exact analytical
and numerical results are available~\cite{RevModPhys.83.1405}.

Among the simplest out-of-equilibrium situation is the dynamics of a
gas after a sudden change (quench) of the external harmonic potential
in which the atoms are confined.  The resulting coherent,
breathing-like oscillations of the atomic cloud are governed by the
collective excitations of the quantum gas, and serve to identify the
quasicondensate (qBEC) regime~\cite{moritz_exciting_2003} and strongly
correlated phases in 1D~\cite{haller_realization_2009}.  The sum-rule
approach has had a remarkable success in the prediction of the
breathing frequencies at zero
temperature~\cite{menotti_collective_2002, astrakharchik_beyond_2005}.
Nonetheless, numerous fundamental questions remain open. These
questions concern the general mechanism of mass transport in
interacting quantum systems as well as its counterpart in momentum
space.  In particular, the effect of temperature on both the lifetime
and the frequency of the collective mode cannot be accessed in terms
of the standard sum-rule approach.  Moreover, breathing oscillations
of the momentum distribution provide complementary insight on the
dynamics of interacting quantum gases, information which is not yet
investigated experimentally nor accessible theoretically to date.

In this Letter, we study experimentally the breathing mode of
harmonically confined 1D Bose gases at finite temperature, comparing
the evolution in real and momentum space.  We report the first
observation of a self-reflection mechanism due to repulsive
interaction, expected of a strongly interacting
system~\cite{minguzzi_exact_2005}.  We measure how the dynamics
evolves through the qBEC to ideal Bose gas (IBG) crossover, and
develop two complementary theoretical treatments for the breathing
modes, taking finite-temperature effects into account.  The
combination of an exact short-time expansion of the dynamics and a
long-wavelength hydrodynamic analysis reveals that the
finite-temperature mass transport happens in an isentropic way, and
provides a qualitatively good agreement with the experimental
observations.

\begin{figure}[htb]
\includegraphics{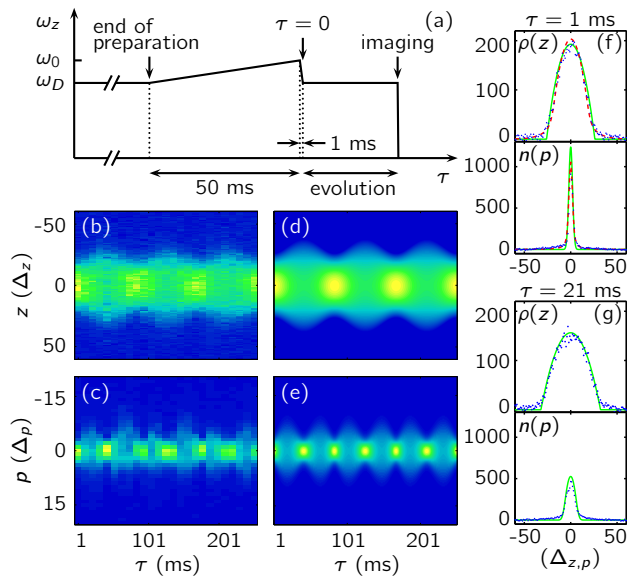}
\caption{\label{fig:intro} (Color online) Quench sequence and
  subsequent density evolution in position and momentum space for a
  qBEC (data A). (a) Longitudinal trapping frequency $\omega_{z}$ as a
  function of time $\tau$.  (b) $\rho(z,\tau)$, (c) $n(p,\tau)$,
  experimental data (in atoms/pixel).  (d), (e), corresponding plots
  from \emph{ab initio} scaling calculation.  (f) and (g) show
  instantaneous $\rho(z)$, $n(p)$ at $\tau=1$ and
  $21$~ms. Experimental data (dots) are compared with the scaling
  solutions (solid lines).  QMC calculation of $n(p,\tau)$ (dashed
  line) is shown in (f).  $z$ ($p$) is in units of pixels $\Delta_{z}$
  ($\Delta_{p}$), with $\Delta_{z}=2.7~\mu$m, and
  $\Delta_{p}=0.14~\hbar/\mu$m. }
\end{figure}

\paragraph{System and quench protocol.}
The Lieb-Liniger (LL) model of 1D bosons in continuum, with a
Hamiltonian
$H_{\text{LL}}=\sum_{j}\frac{p_{j}^{2}}{2m}+\sum_{j<k}g_{\rm
  1D}\delta(z_{j}-z_{k})$, where $m$ is the particle mass, and $g_{\rm
  1D}$($>\!0$) is the coupling constant, is realized experimentally
for gases confined in elongated traps, provided the transverse degrees
of freedom are frozen out~\cite{olshanii_atomic_1998}. Various
experimental probes have been developed for such gases, permitting not
only thermometry methods~\cite{van_amerongen_yang-yang_2008,
  armijo_probing_2010, jacqmin_sub-poissonian_2011,
  vogler_thermodynamics_2013, armijo_mapping_2011,
  jacqmin_momentum_2012}, but also the study of nonequilibrium
behavior.  Of special importance is the measurement of the momentum
distribution, now feasible via Bragg
spectroscopy~\cite{fabbri_momentum-resolved_2011} or focusing
technique~\cite{jacqmin_momentum_2012}.

Our experiment prepares a single tube of $^{87}$Rb gas using an
atom-chip setup~\cite{jacqmin_sub-poissonian_2011}.  The final samples
typically constitute $800$ to $8000$ atoms for this experiment.  The
\emph{in situ} density profile~\cite{van_amerongen_yang-yang_2008}
indicates a temperature around $100$~nK, corresponding to a central
chemical potential $\mu_{0}\in[0.04,0.5]\times k_{B}T$~\footnote{We
  remark that other thermometries are
  possible~\cite{armijo_mapping_2011}, and seem to indicate a lack of
  true equilibrium.  See the discussions following Eq.~(\ref{eqn:HE})
  for more details.  }.  Since the transverse confinement
$\omega_{\perp}=2\pi\times2$ kHz, we have $\mu_{0}\ll
k_{B}T\simeq\hbar\omega_{\perp}$, achieving a nearly 1D scenario. The
breathing mode is excited by quenching the axial confinement
$\omega_{z}$ from $\omega_0$ to $\omega_D$, see
Fig.~\ref{fig:intro}(a) for a sketch.  We keep the quench strength
$\alpha\equiv\omega_{0}/\omega_{D}\simeq1.3$ constant.  The resulting
cloud evolves for a duration $\tau$ before an absorption image is
taken, either \emph{in situ} to yield the density profile
$\rho(z,\tau)$, or at focus to yield the momentum distribution
$n(p,\tau)$. By varying $\tau$, we map out the evolution in real and
momentum space.

\begin{figure*}[hbt]
\includegraphics{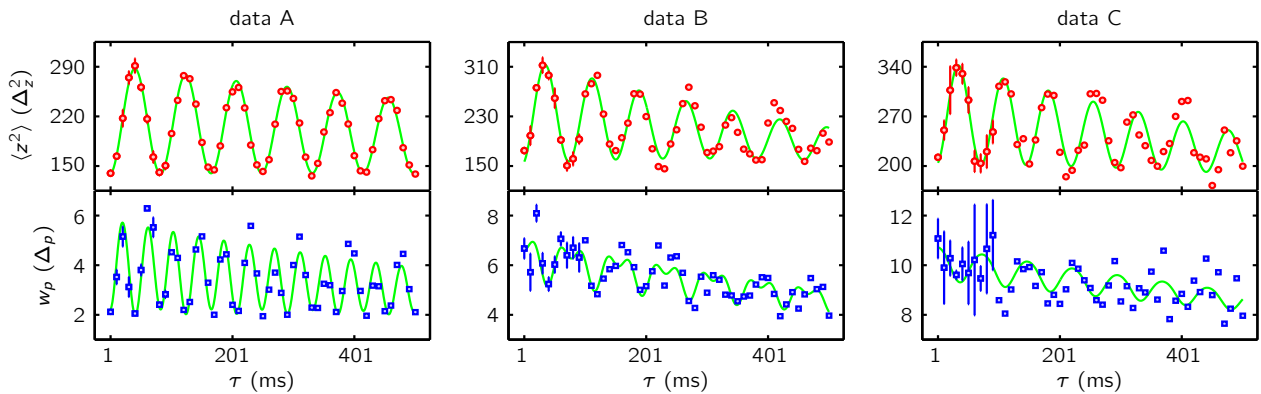}
\caption{\label{fig:osc} Time evolution of the \emph{in situ}
mean-square width $\langle z^2\rangle$ (top) and the momentum HWHM
$w_p$ (bottom) for 3 data sets. The solid lines show the fit: a
damped sinusoid in real space (top), and a two-harmonic model
according to Eq.~(\ref{eqn:fit}) in momentum space (bottom).
Statistical error of the widths are shown for the first $100$~ms.
Data A-C differ in thermodynamic regime.  See Fig.~\ref{fig:cr} and
text for details.  }
\end{figure*}

\paragraph{Quasicondensate regime.}  
We start by considering the qBEC regime: Figures \ref{fig:intro}(b)
and (c) show $\rho(z,\tau)$ and $n(p,\tau)$ for a data set that lies
in this regime (data A).  The time evolution of these two quantities
reveals a remarkable frequency mismatch.  In order to elucidate its
origin, we use the hydrodynamic equations (HDE)~\footnote{The validity of HDE 
for  1D Bose gases at finite temperature, for which a thermalisation time is 
unclear, is not established. Agreement with 
a different theoretical approach (see later in the text) strenghtened this 
hypothesis.}
\begin{eqnarray} 
\partial_{\tau}\rho + \partial_{z}(\rho v) & = & 0,\label{eqn:HDE}\\
\partial_{\tau}v + v\partial_{z}v & = &
- \partial_{z}(\mbox{\ensuremath{\frac{1}{2}}}\omega_{D}^{2}z^{2}) -
\mbox{\ensuremath{\frac{1}{m\rho}}}\partial_{z}P, \nonumber
\end{eqnarray}
where $\rho=\rho(z,\tau)$ and $v=v(z,\tau)$ are the density and
velocity fields, and $P$ is the pressure. For a qBEC, $P=g_{\rm
  1D}\rho^{2}/2$, so that the scaling solution
$\rho(z,\tau)=\frac{1}{b(\tau)}\rho_{0}\big(\frac{z}{b(\tau)}\big)$~\cite{castin_bose-einstein_1996}
is valid provided $\rho_{0}$ is the steady state inverted parabola,
and the scaling factor $b$ obeys
$\ddot{b}+\omega_{z}(\tau)^{2}b=\frac{\omega_{0}^{2}}{b^{2}}$.  Such
$\rho(z,\tau)$ is periodic in time with a frequency $\omega_{Bz}\simeq
\sqrt{3}\omega_{D}$.  Neglecting thermal fluctuations, it follows that
the momentum distribution satisfies~\cite{castin_bose-einstein_1996}
$n(p,\tau)\propto\rho_0(\frac{bp}{\dot{b}m})$, which is periodic in
time with a frequency $\omega_{Bp}=2\omega_{Bz}$. 
Minimal momentum width occurs when the real-space distribution is 
both  the largest and the thinnest. 
The latter scenario, which does not occur for a non-interacting gas, 
corresponds to a self-reflection of the could due to repulsive interactions.
The frequency doubling ($\omega_{Bp}=2\omega_{Bz}$)  is thus a direct 
consequence of atomic interactions. 
In higher dimensions, the scaling
solutions also predicts a frequency doubling of the breathing
modes~\footnote{Both the compression mode and the surface mode for an
  axially symmetric trap.  }.  What may have prevented its observation
is perhaps the fact that early experiments using time-of-flight (TOF)
techniques do not measure the true momentum distribution due to the
contribution of interaction energy.

It is worth mentioning that such a momentum-space frequency doubling
in the oscillatory behavior of $w_p$, the half width at half maximum
(HWHM) in momentum space, is expected to occur for a Tonks gas
($g_{\rm 1D} \rightarrow +\infty$)~\cite{girardeau_relationship_1960}~:
the momentum distribution for large $\alpha$ oscillates between a Fermi-like and 
a Bose-like distribution, a thin Bose-like distribution occuring 
both when the cloud is the thinnest and the largest~\cite{minguzzi_exact_2005}, 
while in the expansion and compression phases the momentum distribution is 
dominated by the large Fermi-like hydrodynamic component.  
This marks the
difference between the breathing behaviour of a strongly interacting
1D Bose gas and that of a noninteracting gas, even though the breathing 
frequency in real
space is $\omega_{Bz}=2\omega_D$ for both cases.  Similar collective
oscillations of strongly interacting 1D Bose gases have been studied
in two experiments~\cite{kinoshita_quantum_2006,
  haller_realization_2009} to our knowledge, neither reporting any
frequency doubling. In the former~\cite{kinoshita_quantum_2006}, a
different excitation scheme is employed, potentially rendering the
frequency doubling inaccessible. In the
latter~\cite{haller_realization_2009}, the finite TOF may have
prevented the direct measurement of the momentum distribution.

In our experiment the momentum-space frequency doubling is clearly
seen in Fig.~\ref{fig:intro}(b-c). Moreover, the data are in good
agreement with the zero-temperature theory above, whose predicted
real- and momentum-space density evolutions for a system initially at
equilibrium in the trap of frequency $\omega_{0}$ are shown in
Fig.~\ref{fig:intro}(d-e), with $n(p,\tau)$ broadened to account for
finite resolution~\footnote{We find a Gaussian point spread function
of about $2$ pixels in rms width sufficient to account for the
broadening.}.  For a more quantitative comparison, we show two
frames of the instantaneous densities.  At $\tau=1$~ms (minimal
\emph{in situ} width, Fig.~\ref{fig:intro}(f)), the scaling solution
predicts a vanishing momentum width, but thermal fluctuations dominate
such that the finite-temperature momentum distribution computed using
quantum Monte Carlo (QMC) methods~\cite{jacqmin_momentum_2012} (dashed
lines) accounts for the data better than the sole effect of
resolution~\footnote{assuming a system at thermal equilibrium in a
  trap $\omega_z=\omega_0$ and at the temperature obtained from
  independent calibration without the $\omega_z$ ramp.}.  However,
thermal fluctuations are small corrections at $\tau=21$~ms (maximal
momentum width, Fig.~\ref{fig:intro}(g)), and both $\rho(z)$ and
$n(p)$ are in excellent agreement with the inverted parabolas of the
scaling solution. Self-similarity of the shapes 
is due to the fact the momentum distribution is dominated by the 
hydrodynamic velocity field, which is linear in position. 
The same phenomena is at the origine of the dynamic fermionization
for a strongly interacting system~\cite{minguzzi_exact_2005}.

\paragraph{Ideal Bose gas regime.} 
For an IBG, a single-particle description suffices and breathing
amounts to a rotation of the phase-space density, so that the widths
in real and momentum space oscillate out of phase at the same
frequency $\omega_{B}=2\omega_{D}$.  Fig.~\ref{fig:osc} shows a data
set close to this regime (data C), where the \emph{in situ}
mean-square width $\langle z^2 \rangle$ (momentum HWHM $w_p$) is
obtained from Gaussian (Lorentzian) fit. The antiphase is apparent
from the plots.  Fitting both time evolution with damped sinusoids, we
measure identical breathing frequency, $\omega_{B}/\omega_{D} = 1.84
\pm 0.04$, where $\omega_{D}$ is determined by monitoring the
center-of-mass (dipole) oscillations.

\begin{figure}[bt]
\includegraphics{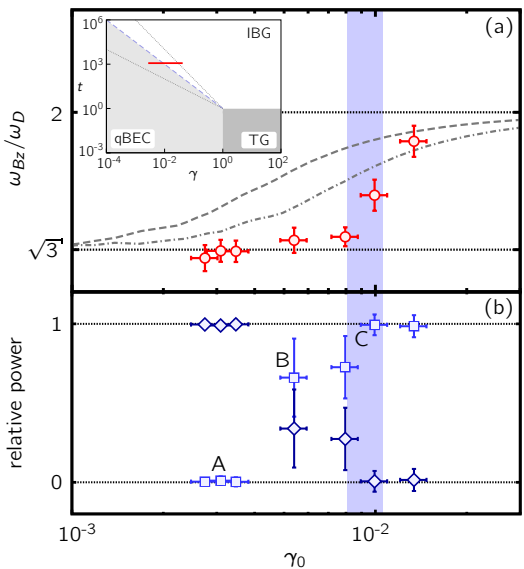}
\caption{\label{fig:cr} Breathing mode through the crossover.  (a)
  breathing frequency in units of dipole frequency, as a function of
  $\gamma_{0}$.  Our theory evaluated at $t=1100$ (dashed line) and
  $t=400$ (dash-dotted line) are shown.  (b) relative power of the
  first (square) and second (diamond) harmonic of $w_p$.  Data labels
  (A-C) correspond to those in Fig.~\ref{fig:osc}.  The errorbars
  account for fitting error only.  The dotted lines are the asymptotic
  limits.  The shaded region shows the crossover criteria
  $\gamma_{co}=t^{-2/3}$ (the dashed line in the inset), at $t\simeq
  1100$ given by the \emph{in situ} profile with $20\%$
  uncertainly. Inset: the phase diagram of the LL
  model~\cite{kheruntsyan_finite-temperature_2005}, where all lines
  represent smooth crossovers.  The horizontal line segment shows the
  region explored by the data~\cite{jacqmin_momentum_2012}.  }
\end{figure}

\paragraph{Crossover in real space.}
A central point to be understood is how the breathing frequency varies
through the qBEC to IBG crossover. 
To address
this question, we vary the total atom number in order to explore
different regimes.  The inset of Fig.~\ref{fig:cr} shows the region
spanned by the data in the ($\gamma$, $t$) phase diagram of the LL
model~\cite{kheruntsyan_finite-temperature_2005}, where
$\gamma=\frac{mg_{\rm 1D}}{\hbar^{2}\rho}$ is the (local) interaction
parameter, and $t=\frac{2\hbar^{2}k_{B}T}{mg_{\rm 1D}^{2}}$ is the
reduced temperature. Each sample is characterized by $\gamma_{0}$,
evaluated at the peak density. We extract $\langle z^2\rangle$ from
fitting $\rho(z)$ with either an inverted parabola for
$\gamma_{0}<0.004$~\footnote{The division is heuristically justified
  by the increasingly significant wings for greater values of
  $\gamma_0$.  } or a Gaussian otherwise. We obtain the real-space
breathing frequency $\omega_{Bz}$ by fitting $\langle
z^2\rangle(\tau)$ with a damped sinusoid.  The measured
$\omega_{Bz}/\omega_{D}$ as a function of $\gamma_{0}$, shown in
Fig.~\ref{fig:cr}(a), displays a smooth crossover between the
asymptotic theoretical limits $\sqrt{3}$ and
2~\cite{menotti_collective_2002}.  Such a frequency shift was first
observed in \cite{moritz_exciting_2003}, where the authors probed the
breathing behaviour of an ensemble of 1D gases in real space.  In
addition to the issue of thermalization mentioned in
\cite{moritz_exciting_2003}, the intrinsic ensemble averaging renders
it problematic to characterize the system with a single value of
temperature $T$ or interaction parameter $\gamma$.  In contrast, our
current experiment with a single 1D system allows for a more
quantitative study of the problem at hand.

In order to provide a theoretical treatment of the crossover, we
devise two complementary approaches.  On one hand, we model the
crossover using the HDE Eq.~(\ref{eqn:HDE}).  Since long-wavelength
density waves in a fluid are adiabatic~\footnote{The two-fluid model
  that predicts a second sound in higher dimensions do not apply.  },
we use the isentropic pressure curves derived numerically from the YY
EoS~\cite{yang_thermodynamics_1969}.  The breathing mode frequency
measured experimentally does not depend on the oscillation amplitude
for the explored parameter range.  We thus linearize the HDE for small
displacement and $\omega_{Bz}$ is obtained by solving an eigenvalue
problem.  Results evaluated at $t=1100$ are shown as a dashed line in
Fig.~\ref{fig:cr}(a).  On the other hand, we provide a microscopic
treatment of the breathing frequency that accounts for the effect of
temperature.  Assuming the system at $\tau=0$ is at thermal
equilibrium with a Hamiltonian $H = H_{\text{LL}}+H_{\text{pot}}$,
$H_{\text{pot}}=m\omega_{0}^{2}\sum_{j}z_{j}^{2}/2$, the expansion of
the Heisenberg equation of motion after a quench $\Delta
H=m(\omega_{D}^{2}-\omega_{0}^{2})\sum_{j}z_{j}^{2}/2$ gives $\langle
\Delta H \rangle (\tau) = \langle \Delta H \rangle_{T} -
\mbox{\ensuremath{\frac{\tau^{2}}{2}}} \langle C_{2} \rangle_{T} +
\mbox{\ensuremath{\frac{\tau^{4}}{4!}}}  \langle C_{4} \rangle_{T} +
\hdots$, where $C_{2}$ and $C_{4}$ are the second and fourth order
nested commutators with $H_{f}=H+\Delta H$, and the thermal average
$\langle~\rangle_{T}$ is taken over the thermal state of $H$. Suppose
the time evolution $\langle\Delta H\rangle(\tau)\propto\langle
z^{2}\rangle(\tau)$ is purely sinusoidal at the frequency
$\omega_{Bz}$, we have
\begin{equation} 
\frac{\omega_{Bz}^{2}}{\omega_{D}^{2}}=\frac{\langle
  C_{4}\rangle_{T}}{\langle C_{2}\rangle_{T}}=4-\frac{1}{2}
\frac{\langle H_{{\rm \text{int}}}\rangle_{T}}{\langle
  H_{\text{pot}}\rangle_{T}},\label{eqn:HE}
\end{equation}
where $H_{\text{int}}=g_{\rm 1D}\sum_{j<k}\delta(z_{j}-z_{k})$ is the
interaction part of the Hamiltonian.  The final equality comes from
explicitly working out $\langle C_{2}\rangle_{T}$ and $\langle
C_{4}\rangle_{T}$, and taking the limit of an infinitesimal quench
amplitude $\alpha \rightarrow 1$, where the sinusoidal approximation
is valid~\footnote{Eq.~(\ref{eqn:HE}) could be used for a quench of
  finite amplitude ($\alpha\neq1$) \emph{a priori}.  However, in the
  qBEC regime, a comparison with the scaling solution indicates that
  the sinusoidal approximation breaks down at finite $\alpha$}.

It is remarkable that both approaches (HDE and the exact short-time
expansion) give the same results for the breathing frequencies through
the crossover.  This indicates that the finite-temperature
mass transport is indeed isentropic.  Moreover, Eq.~(\ref{eqn:HE})
extends the regime of validity of the existing sum-rule approaches,
which erroneously predicts $\omega_{Bz}=\sqrt{2}\omega_{D}$ in IBG,
whereas the true asymptotic limit is $\omega_{Bz}=2\omega_{D}$.

Our theories provide a qualitatively good agreement with the
experimental data, despite an overall overestimation of the breathing
frequency. This quantitative departure most likely indicates that the
atomic ensemble is not at thermal equilibrium at the end of the
preparation (evaporation).  Although the \emph{in situ} profiles
calibrated with the YY EoS generally yields about $100$~nK, the
atom-number fluctuation indicates approximately $35$ to $45$~nK at the
center of the cloud.  We thus include in Fig.~\ref{fig:cr}(a) our
prediction at $t=400$ (dash-dotted line).  We believe that such a lack
of equilibrium is a direct consequence of evaporating into the
integrable (1D) regime.  Further investigation in this direction is
underway.

\paragraph{Disappearance of the self-reflection mechanism.} 
To address the behavior of the breathing mode in momentum space, we
investigate the time evolution of $w_p$, extracted from a Lorentzian
fit of $n(p)$. Since $w_p$ shows a periodic behavior at the frequency
$\omega_{Bz}$, it can be expanded in a discrete Fourier spectrum.  The
relative weight of the Fourier components varies through the
crossover, and we obtain quantitative information by fitting $w_p$
with a function
\begin{equation}\label{eqn:fit}
y = Ae^{-\frac{\tau}{\tau_{1}}} + B e^{-\frac{\tau}{\tau_{2}}} \Big[\sqrt{K}
\cos(\omega_{Bz}\tau) - \sqrt{1-K} \cos(2\omega_{Bz}\tau) \Big],
\end{equation}
shown in Fig.~\ref{fig:osc} (bottom, solid line), with $A$ fixed at
the average width during the initial cycle and the phase corresponds
to a minimal \emph{in situ} width at the start of the oscillations.
The relative power of the first harmonic $K$ (squares) and second
harmonic $1-K$ (diamonds) are shown in Fig.~\ref{fig:cr}(b) as a
function of $\gamma_{0}$.  In qBEC, the second harmonic dominates, as
predicted by the scaling solution, and signals the self-reflection
mechanism.  In IBG, the first harmonic dominates, as expected for a
noninteracting gas where the self-reflection is absent.  Both weights
vary gradually through the crossover, indicating a smooth
disappearance of the self-reflection mechanism.  This can be seen as
the effect the breathing mode has on the thermally excited Bogoliubov
modes of high energy, e.g.~their frequency and wavefunction would be
modulated in time, such that $w_p$ is larger at minimal $\langle
z^2\rangle$ than that at maximal $\langle z^2\rangle$, see
Fig.~\ref{fig:osc} (data B).  The periodicity at $2\omega_{Bz}$ is
then broken and the first harmonic at $\omega_{Bz}$ emerges.
Figure~\ref{fig:cr} shows that the first harmonic starts to gain
weight at a value of $\gamma_{0}$ significantly smaller than that
where the frequency shift takes place in real space.

The breathing mode observed has lifetime estimated to be on the order
of seconds.  This is in stark contrast with situations in 3D where
damping of the collective modes occurs, mainly via Landau damping
mechanism~\cite{jin_temperature-dependent_1997,fedichev_damping_1998}.
The long lifetime of the breathing mode in 1D may be related to the
integrability of the underlying LL model.

\paragraph{Conclusions.} 
In this Letter we have probed the breathing mode in real and momentum
space through the qBEC to IBG crossover.  The shift of the real-space
frequency between the asymptotic values $\sqrt{3}\omega_D$ and
$2\omega_D$ is demonstrated.  Our theory models that assume thermal
equilibrium before the quench do not agree with the measurement
quantitatively, indicating the possibility of a non-Gibbs initial
state produced by evaporation.  We report the first observation of a
momentum-space frequency doubling in qBEC, corresponding to an
interaction-induced self-reflection mechanism that is expected of a
Tonks gas.  We experimentally map out the disappearance of the
self-reflection through the crossover, for which no theoretical
predictions exist to date. This illustrates the richness of
out-of-equilibrium dynamics and the importance of the momentum-space
observation to unveil the underlying physics.


\begin{acknowledgments}
The authors would like to thank A.\ Minguzzi, T.\ Roscilde, S.\ Stringari,
P.\ Vignolo, M.\ Zvonarev for stimulating discussions.  This work is
financially supported by Cnano IdF, the Austro-French FWR-ANR Project
I607, and the FP7-Marie Curie IEF grant 327143.
\end{acknowledgments}

\bibliographystyle{apsrev}

\end{document}